\documentclass[a4paper]{jpconf}
\usepackage{graphicx}
\usepackage{amsmath}
\usepackage{amsfonts}
\begin{document}
\title{New solution of the $\mathcal{N}=2$ Supersymmetric KdV equation via Hirota methods.}

\author{Laurent Delisle and V\'eronique Hussin}

\address{D\'epartement de math\'ematiques et statistiques, Universit\'e de Montr\'eal,\\
C.P. 6128, Succc. Centre-ville, Montr\'eal, (QC) H3C 3J7, Canada}

\ead{delisle@dms.umontreal.ca}

\begin{abstract}
We consider the resolution of the $\mathcal{N}=2$ supersymmetric KdV equation with $a=-2$ ($SKdV_{a=-2}$) from the Hirota formalism. For the first time, a bilinear form of the $SKdV_{a=-2}$ equation is constructed. We construct multisoliton solutions and rational similarity solutions.
\end{abstract}

\section{Introduction}
Let us recall that $\mathcal{N}=2$ supersymmetric (SUSY) extensions of the KdV equation have been largely studied in the past \cite{LM,AHW,IMM,GS,HK} in terms of integrability conditions and different types of solutions. Such extensions are given as a one-parameter ($a\in \mathbb{R}$) family of Grassmann-valued partial differential equations with one dependent variable $A(x, t;\theta_{1}, \theta_{2})$ where the independent variables are given as a set of even (commuting) space $x$ and time $t$ variables and a set of odd (anticommuting) variables $\theta_{1}$, $\theta_{2}$. The dependent variable $A$ is supposed to be an even superfield and satisfies 
\begin{equation}
 A_{t}=(-A_{xx}+(a+2) AD_{1}D_{2}A+(a-1)(D_{1}A)(D_{2}A)+a A^{3})_{x},
 \label{N2mkdv}
\end{equation}
 where $D_{1}$, $D_{2}$ are the covariant superderivatives given by
\begin{equation}
 D_{i}=\theta_{i}\partial_{x}+\partial_{\theta_{i}},\quad i=1,2,
\end{equation}
and such that $D_{1}^2=D_{2}^2=\partial_{x}$. Note that since $D_{i}^{2}=\partial_{x}$ ($i=1,2$), $D_{i}$ can be viewed as the square root of $\partial_{x}$. In \cite{LM}, it was shown that equation (\ref{N2mkdv}) is integrable for the special values $a=-2,1,4$.

Since the odd variables satisfy $\theta_{1}^2=\theta_{2}^2=\theta_{1}\theta_{2}+\theta_{2}\theta_{1}=0$, the dependent variable $A$ admits the following Taylor expansion 
\begin{equation}
 A(x,t;\theta_{1},\theta_{2})=u(x,t)+\theta_{1}\xi_{1}(x,t)+\theta_{2}\xi_{2}(x,t)+\theta_{1}\theta_{2}v(x,t),
 \label{decompA}
\end{equation}
where $u$ and $v$ are commuting (bosonic) complex valued functions and $\xi_{1}$ and  $\xi_{2}$ are anticommuting (fermionic) complex valued functions.

Solutions of equation (\ref{N2mkdv}) have been obtained using different methods. Among them, an original approach \cite{AHW} has adapted the classical symmetry reduction method to the SUSY context. Indeed in this approach, starting with the SUSY equation (\ref{N2mkdv}), it has been reduced to a set of only one bosonic and one fermionic equations using invariance properties associated with invariant superalgebras. 

In the following, we will take $a=-2$ in equation (\ref{N2mkdv}) where it is well-known that we get solitons solutions as travelling wave solutions \cite{AHW,IMM,GS} and also rational similarity solutions as in the classical case \cite{K,DJ}. Such solutions will be generalized in the SUSY case using the Hirota formalism. Note that the Hirota formalism was adapted to the case $a=1$ and $a=4$ \cite{ZLW} but was still an open problem for the case $a=-2$.

Let us write the set of equations we are working with in this case. The equation (\ref{N2mkdv}) becomes: 
\begin{equation}
 A_{t}=(-A_{xx}-3(D_{1}A) (D_{2}A)-2 A^{3})_{x},
 \label{N2mkdva-2}
\end{equation}
and admits the decomposition
 \begin{align}
u_{t}+(u_{xx}+2 u^3+3 \xi_{1}\xi_{2})_x&=0,\label{GetmKdVa-2}\\
v_{t}+(v_{xx}+3v^2+6 u^2v+3 u_{x}^2)_x-(3\xi_{2}\xi_{2; x}+12 u \xi_{1}\xi_{2})_x&=0,\label{GetKdVa-2}\\
\xi_{1; t}+(\xi_{1; xx} +3(v+2 u^2)\xi_{1}+3 u_{x}  \xi_{2} )_x&=0,\label{ksi1a-2}\\
\xi_{2; t}+(\xi_{2; xx} +3(v+2 u^2)\xi_{2}-3 u_{x}  \xi_{1} )_x&=0.\label{ksi2a-2}
\end{align}

It is interesting to note that in the one-fermionic parameter bosonization \cite{ARS} which consists of writing the fermionic fields in (\ref{decompA}) as $\xi_{i}=\zeta f_{i}$ ($i=1,2$), with $f_{i}$ a bosonic complex function and $\zeta$ an odd parameter such that $\zeta^{2}=0$, equations (\ref{GetmKdVa-2}) and (\ref{GetKdVa-2}) decouples into purely bosonic equations. In particular, we see that (\ref{GetmKdVa-2}) reduces to the classical mKdV equation.


\section{Hirota formalism and solutions for the case $a=-2$.}
The Hirota formalism is a well known process in the classical \cite{DJ,AS,AS1} and in the $\mathcal{N}=1$  SUSY cases \cite{MY,C1,CRG,GS1}. This formalism has been used, in particular, to obtain $N$-soliton solutions. We examine the possible generalizations in the $\mathcal{N}=2$ SUSY case following the idea given in \cite{ZLW}.  We write the superfield (\ref{decompA}) in the following form
\begin{equation}
 A(x,t;\theta_{1},\theta_{2})=u^{b}(x,t;\theta_{1})+\theta_{2}\ \xi^{f}(x,t;\theta_{1}),
 \label{AHirota}
\end{equation}
 where $u^{b}, \ \xi^{f}$ are even and odd superfields, respectively. Comparing with  (\ref{decompA}) , we have $u^{b}=u+\theta_{1}\xi_{1}$ and $\xi^{f}= \xi_{2}-\theta_{1} v$.We thus introduce the following change of variable $A=\partial_{x} B$, where 
\begin{equation}
B(x,t;\theta_{1},\theta_{2})={\cal U}^{b}(x,t;\theta_{1})+\theta_{2}\ \eta^{f}(x,t;\theta_{1}), 
\label{be}
\end{equation}
to equalize the order of the equation with the number of appearence of the $x$ derivative in the nonlinear terms. Equation (\ref{N2mkdva-2}) thus becomes, after integrating once,
\begin{equation}
 B_{t}=-B_{xxx}-3(D_{1}B_{x})(D_{2}B_{x})-2B_{x}^{3},
 \label{eqube}
\end{equation}
where the constant of integration is set to zero. Inserting the explicit form (\ref{be}) of $B$ in equation (\ref{eqube}), we get a set of coupled $\mathcal{N}=1$ SUSY equations on ${\cal U}^{b}$ and $\eta^{f}$:
\begin{align}
 {\cal U}^{b}_{t}+{\cal U}^{b}_{xxx}-3\eta^{f}_{x}D_{1}{\cal U}^{b}_{x}+2({\cal U}^{b}_{x})^{3}&=0,
 \label{equv}\\
\eta^{f}_{t}+\eta^{f}_{xxx}-3\eta^{f}_{x}D_{1}\eta^{f}_{x}-3{\cal U}^{b}_{xx}D_{1}{\cal U}^{b}_{x}+6\eta^{f}_{x}
({\cal U}^{b}_{x})^{2}&=0.
\label{equeta}
\end{align}

 Now the standard strategy is to introduce a change of the dependent variables, ${\cal U}^{b}$ and $\eta^{f}$, in such a way that equations (\ref{equv}) and (\ref{equeta}) become quadratic. 
 



Since equation (\ref{equv})  is a modified mKdV equation, we use the change of variables:
\begin{equation}
 {\cal U}^{b}=-i \log(\dfrac{\tau_{1}}{\tau_{2}}), \quad \eta^{f}= D_{1} \log(\dfrac{\tau_{1}}{\tau_{2}}), \label{veta}
\end{equation}
 where $\tau_{1}=\tau_{1}(x,t;\theta_{1})$ and $\tau_{2}=\tau_{2}(x,t;\theta_{1})$ are both bosonic superfields.
 
From (\ref{veta}), we deduce that
\begin{equation}
 \xi^{f}=i D_{1}u^{b},
 \label{xispe}
\end{equation}
which imposes the constraint $v=-i u_{x}$ and $\xi_{2}=i \xi_{1}$ on the original superfield $A=$(\ref{decompA}). The constraints on the field $\xi_{2}$  and $v$ are standard when dealing with chiral superfields \cite{FG}. A chiral superfield $A$ satisfies the constraint $D_{+}A=0$ where $D_{+}=\frac{1}{2}(D_{1}+i D_{2})$. Here, we see that these constraints are imposed in order to produce a bilinear form.

From (\ref{veta}), we see that $\eta^{f}=i D_{1}\mathcal{U}^{b}$ which reduces the pair of equations (\ref{equv}) and (\ref{equeta}) to the mKdV equation
\begin{equation}
 {\cal U}^{b}_{t}+{\cal U}^{b}_{xxx}+2({\cal U}^{b}_{x})^{3}=0.
 \label{mkdvv}
\end{equation}
Let us mention that (\ref{mkdvv}) is still a SUSY equation and using the decomposition $ {\cal U}^{b}= {\cal U}_0^{b}(x,t)+\theta_{1}\varphi(x,t)$, we get the classical mKdV equation for ${\cal U}_0^{b}$ and $\varphi$ satisfies
\begin{equation}
 \varphi_{t}+\varphi_{xxx}+6({\cal U}^{b}_{0; x})^{2}\varphi_{x}=0,
 \label{mkdvphi}
\end{equation}
for which a particular solution is $\varphi=\zeta \ {\cal U}^{b}_{0; x}$ where $\zeta$ is an odd constant. Such a result is common when we deal with ${\cal N}=1$ SUSY KdV and mKdV equations \cite{MY,C1,CRG,GS1}.

We thus have a direct bilinearization of equations  (\ref{equv}) and (\ref{equeta})
\begin{align}
 (\mathcal{D}_{t}+\mathcal{D}_{x}^{3})(\tau_{1}\cdotp\tau_{2})&=0,\label{H1}\\
\mathcal{S}\mathcal{D}_{x}(\tau_{1}\cdotp\tau_{2})&=0\label{H2},
\end{align}
where 
\begin{equation}
\mathcal{S}\mathcal{D}_{x}^{n}(\tau_{1}\cdotp\tau_{2})=(D_{\Theta_{1}}-D_{\Theta_{2}})(\partial_{x_{1}}-\partial_{x_{2}})^{n}
\tau_{1}(x_{1};\Theta_{1})\tau_{2}(x_{2};\Theta_{2})\lvert_{x=x_{1}=x_{2},\theta_{1}=\Theta_{1}=\Theta_{2}},
\end{equation}
is the super Hirota derivative and $D_{\Theta_{i}}=\partial_{\Theta_{i}}+\Theta_{i}\partial_{x_{i}}$, $i=1,2$. Equations (\ref{H1}) and (\ref{H2}) are a natural generalization to the SUSY case of the classical bilinear form of the mKdV equation. 



\subsection{Super soliton solutions.}
The Hirota formalism helps us to recover the travelling wave solution or one super soliton solution  but we also get the $N$ super soliton solutions \cite{GS,ZLW}.
Indeed, for the travelling wave solution, we take 
\begin{equation}
 \tau_{1}=1+a_{1} e^{\Psi}, \quad \tau_{2}=1+b_{1} e^{\Psi},
\end{equation}
where $\Psi=(\kappa x+\omega t)+\theta_{1}\zeta$ and $a_{1}$ and $b_{1}$ are nonzero even parameters. Introducing $\tau_{1}$ and $\tau_{2}$ in equation (\ref{H2}) yields the following relation
\begin{equation}
 b_{1}=-a_{1}.
\end{equation}
The dispersion relation
\begin{equation}
 \omega+\kappa^{3}=0,
\end{equation}
is obtained from the equation (\ref{H1}) from which we get  $\Psi=\kappa (x-\kappa^2 t)+\theta_{1}\zeta$. It is expected since we are interested in travelling wave solutions.  Since 
\begin{equation}
 {\cal U}^{b}=F(\Psi)= F(\Psi_0) +\theta_{1}\zeta \frac{dF}{d\Psi}|_{\Psi=\Psi_0},
 \label{Ube}
\end{equation}
where $\Psi_0=\kappa(x-\kappa^2 t)$, and $u+\theta_{1}\xi_{1}=\partial_{x}{\cal U}^{b}$, we see that the fermionic solution $\xi_1$ is essentially the derivative with respect to $x$ of $u$.

We easily recover a one super soliton choosing $a_{1}=i$ and $\kappa=1$. In such a case, the $u$ component of the original superfield $A$ is the classical soliton solution of the mKdV equation given by $u(x,t)={\text{sech}}(x-t)$.


Now let us exhibit new solutions known as $N$ super soliton solutions \cite{GS,HK,MY,C1,CRG,GS1}.




Indeed, in order to find the $2$ super soliton solution, we first take
\begin{align}
 \tau_{1}&=1+a_{1}e^{\Psi_{1}}+a_{2}e^{\Psi_{2}}+a_{1}a_{2}A_{12}e^{\Psi_{1}+\Psi_{2}},\\
\tau_{2}&=1+b_{1}e^{\Psi_{1}}+b_{2}e^{\Psi_{2}}+b_{1}b_{2}B_{12}e^{\Psi_{1}+\Psi_{2}},
\end{align}
where now $\Psi_{i}=\kappa_{i}x+\omega_{i}t+\theta_{1}\zeta_{i}$ and $a_{i}$ and $b_{i}$ are nonzero even parameters ($i=1,2$). This type of solution as for effect of breacking translation invariance. Introducing these expressions in the mKdV bilinear form (\ref{H1}) and (\ref{H2}), the first equation yields the expected dispersion relations:
\begin{equation}
 \omega_{i}+\kappa_{i}^{3}=0, \quad i=1,2.
\end{equation}
The second equation gives the following conditions:
\begin{align}
 b_{i}&=-a_{i}, \ i=1,2,\\
A_{12}&=B_{12}=\left(\dfrac{\kappa_{1}-\kappa_{2}}{\kappa_{1}+\kappa_{2}}\right)^{2}
\end{align}
and the standard condition \cite{GS,C1} relating the anticommuting variables $\zeta_{1}$ and $\zeta_{2}$ given by ($\kappa_{1}, \kappa_{2}\neq 0$)
\begin{equation}
 \kappa_{1}\zeta_{2}=\kappa_{2}\zeta_{1}.
\end{equation}
Finally, the $\tau$-functions are given by
\begin{align}
 \tau_{1}&=1+a_{1}e^{\Psi_{1}}+a_{2}e^{\Psi_{2}}+a_{1}a_{2}A_{12}e^{\Psi_{1}+\Psi_{2}},\\
\tau_{2}&=1-a_{1}e^{\Psi_{1}}-a_{2}e^{\Psi_{2}}+a_{1}a_{2}A_{12}e^{\Psi_{1}+\Psi_{2}}.
\end{align}
We may enjoy the behavior of the $u(x,t)$ part of the $2$-soliton solution
\begin{align}
 u(x,t)=
-\dfrac{9(10\cosh(\frac{1}{8}(t-4x))+5\cosh(t-x)+8\sinh(\frac{1}{8}(t-4x))+4\sinh(t-x))}{72+41\cosh(\frac{3}{8}(3t-4x))+81\cosh(\frac{1}{8}(7t-4x))+40\sinh(\frac{3}{8}(3t-4x))},\label{grosse}
\end{align}
 where the parameters $\kappa_{1}$, $\kappa_{2}$, $a_{1}$, $a_{2}$ are chosen as $\kappa_{1}=2\kappa_{2}=1$ and $a_{1}=a_{2}=i$, so that $\Psi_{1}=(x-t)+\theta_{1}\zeta_{1}$ and $\Psi_{2}=\frac{1}{8}(4x-t)+\theta_{1}\zeta_{2}$. In Figure 1, we give the solution $|u|$ and the fermionic part $-f_{1}$ in $\xi_{1}=\zeta f_{1}=\zeta u_{x}$ for $t=-10, 0, 10$.

\begin{figure}[!h]
\begin{center}
 \includegraphics[width=4.5cm]{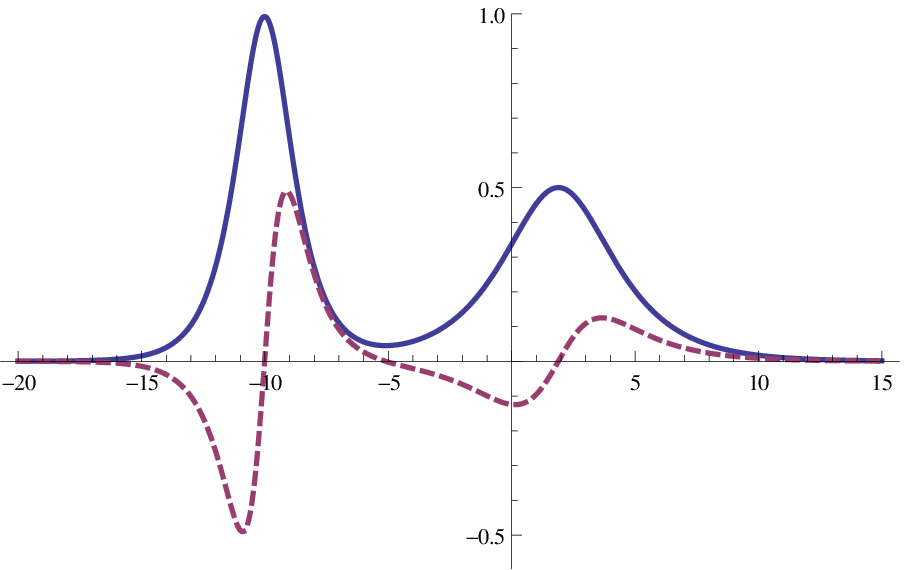}
\includegraphics[width=4.5cm]{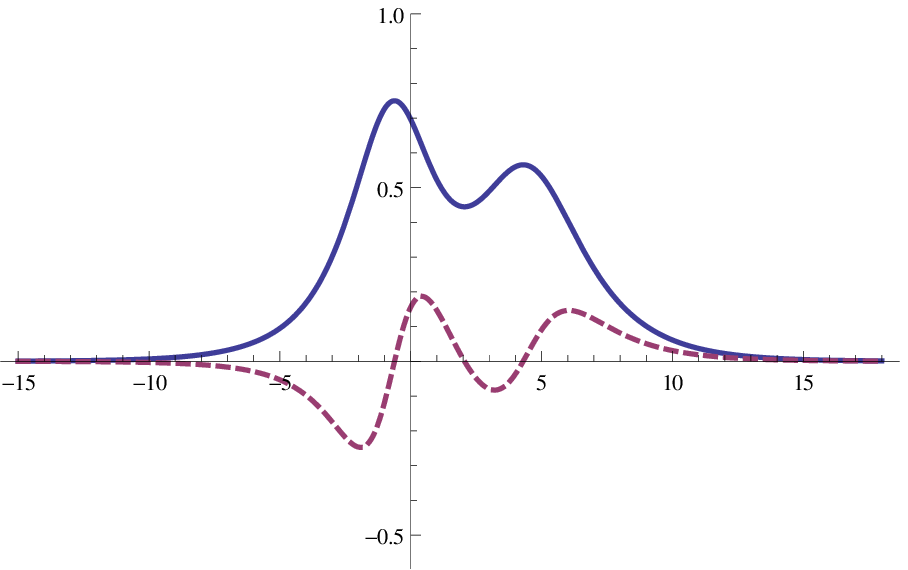}
\includegraphics[width=4.5cm]{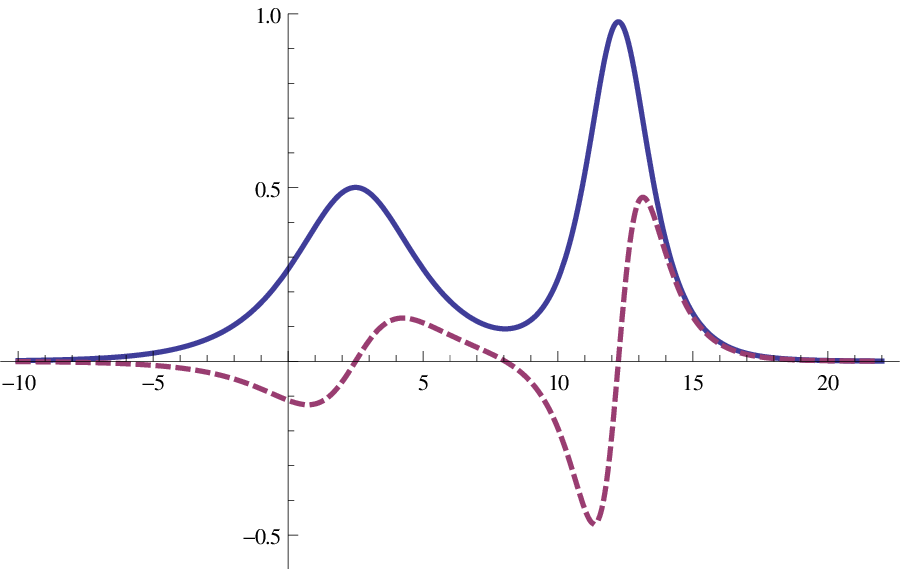}
\caption{The functions $|u(x,t)|$ (full curves) and $-f_{1}(x,t)$ (dashed curves) for $t=-10, 0, 10$.}
\end{center}
\end{figure}

 For the three super soliton solution, $\tau_{1}$ and $\tau_{2}$ take the explicit form:
\begin{align}
 \tau_{1}=1+a_{1}e^{\Psi_{1}}+a_{2}e^{\Psi_{2}}+a_{3}e^{\Psi_{3}}+a_{1}a_{2}A_{12}e^{\Psi_{1}+\Psi_{2}}+a_{1}a_{3}A_{13}e^{\Psi_{1}+\Psi_{3}}\notag\\
+a_{2}a_{3}A_{23}e^{\Psi_{2}+\Psi_{3}}+a_{1}a_{2}a_{3}A_{12}A_{13}A_{23}e^{\Psi_{1}+\Psi_{2}+\Psi_{3}},\\
\tau_{2}=1-a_{1}e^{\Psi_{1}}-a_{2}e^{\Psi_{2}}-a_{3}e^{\Psi_{3}}+a_{1}a_{2}A_{12}e^{\Psi_{1}+\Psi_{2}}+a_{1}a_{3}A_{13}e^{\Psi_{1}+\Psi_{3}}\notag\\
+a_{2}a_{3}A_{23}e^{\Psi_{2}+\Psi_{3}}-a_{1}a_{2}a_{3}A_{12}A_{13}A_{23}e^{\Psi_{1}+\Psi_{2}+\Psi_{3}},
\end{align}
where $\Psi_{i}=\kappa_{i}x-\kappa_{i}^{3}t+\theta_{1}\zeta_{i}$ ($i=1,2,3$) and \cite{GS,C1}
\begin{align}
 A_{ij}&=\left(\dfrac{\kappa_{i}-\kappa_{j}}{\kappa_{i}+\kappa_{j}}\right)^{2},\\
\kappa_{i}\zeta_{j}&=\kappa_{j}\zeta_{i},
\end{align}
for $i,j=1,2,3$ ($i\neq j$). We may enjoy the behavior of the $u(x,t)$ part of the $3$-soliton solution. In Figure 2, we show again $|u|$ and the fermionic part $-f_{1}$ in $\xi_{1}=\zeta f_{1}=\zeta u_{x}$ for $t=-15, 0, 15$ . The parameters $\kappa_{1}$, $\kappa_{2}$, $\kappa_{3}$, $a_{1}$, $a_{2}$, $a_{3}$ are chosen as $\kappa_{1}=\frac{10}{7}\kappa_{2}=\frac{5}{2}\kappa_{3}=1$ and $a_{1}=a_{2}=a_{3}=i$.

\begin{figure}[!h]
\begin{center}
 \includegraphics[width=4.5cm]{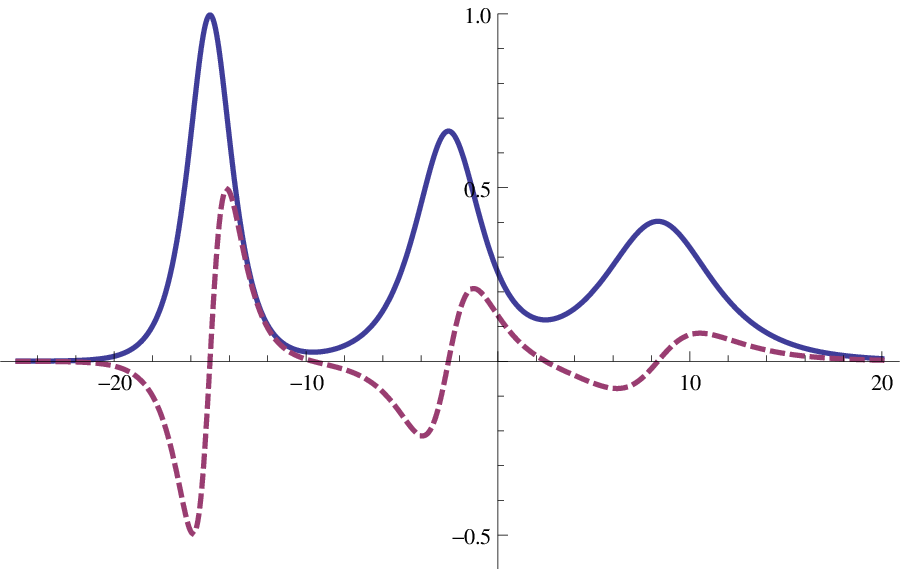}
\includegraphics[width=4.5cm]{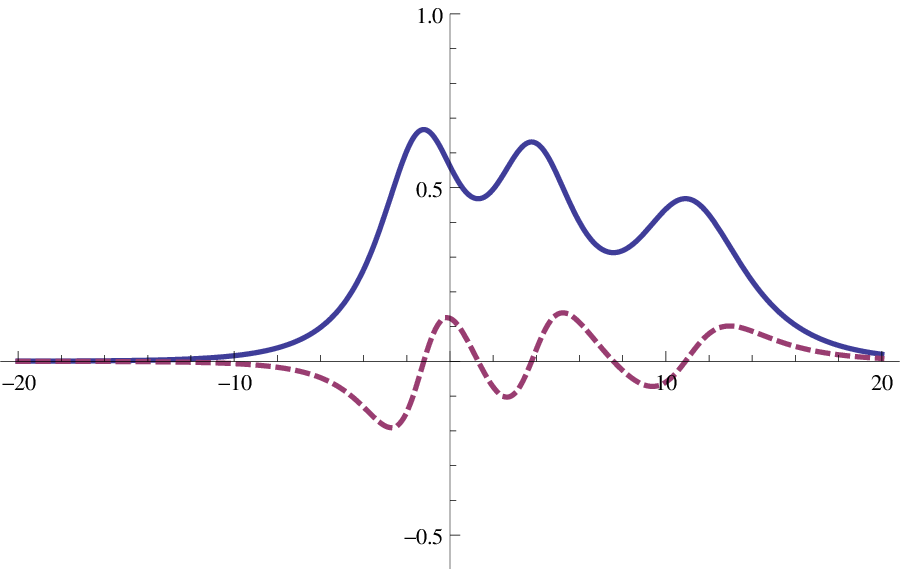}
\includegraphics[width=4.5cm]{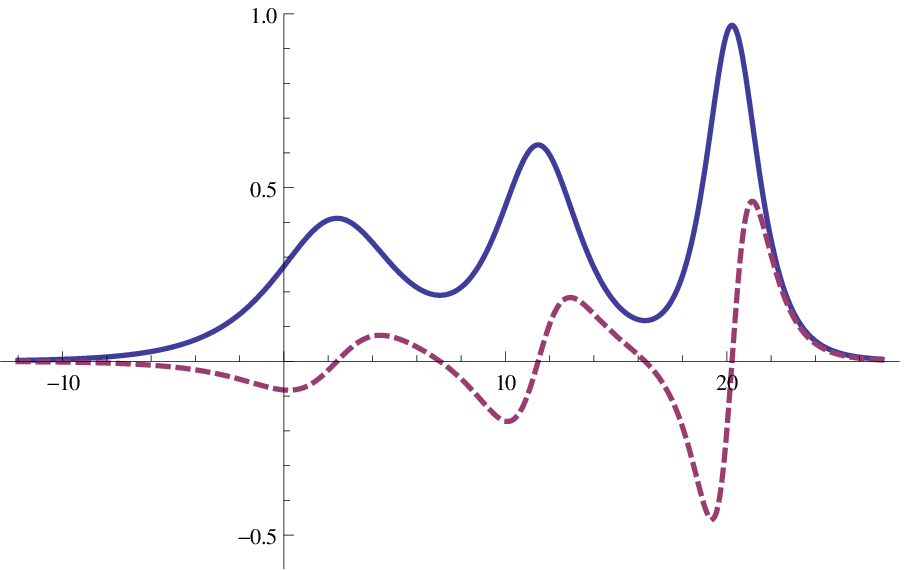}
\caption{The functions $|u(x,t)|$ (full curves) and $-f_{1}(x,t)$ (dashed curves) for $t=-15, 0, 15$.}
\end{center}
\end{figure}

 We have thus shown how to construct the $N$ super soliton solution ($N=1,2,3$), using the bilinear form (\ref{H1}) and (\ref{H2}), by giving the explicit forms of the functions $\tau_{1}$ and $\tau_{2}$. The $N$ super soliton solution ($N\geq3$) is easily generalized using the constraints above \cite{GS,C1,AS1}. We are presently working on a \textit{Mathematica} program which generates the $N$ soliton solution by constructing the $\tau$-functions $\tau_{1}$ and $\tau_{2}$ for general $N$.

\subsection{Rational similarity solutions.}

The Hirota formalism can also be used to get rational similarity solutions \cite{AS}. We assume now a SUSY generalization where the dependent variables $ \tau_{1}$ and  $\tau_{2}$ are polynomials in the independent variable ${\tilde z}= t^{-\frac13}(x+\theta_{1} \zeta)$ and time $t$. Note that ${\tilde z}\lvert_{\theta_{1}=0}=t^{-\frac13}x$ which is the invariant independent variable of the classical mKdV equation under dilatation \cite{DJ}. For example, we take
\begin{equation}
 \tau_{1}(x,t;\theta_{1})= \tau_{1}({\tilde z} ,t)= t^{\frac{1}{3}}{\tilde z}, \quad  \tau_{2}(x,t;\theta_{1})= \tau_{2}({\tilde z} ,t)=t({\tilde z}^{3}+12),
 \label{poly1}
\end{equation}
which solves the bilinear system (\ref{H1}) and (\ref{H2}). We thus have the following form 
\begin{equation}
{\cal U}^{b}(x,t;\theta_{1})={\cal U}^{b}({\tilde z} ,t)= i \log\left(t^{\frac{2}{3}}\dfrac{{\tilde z}^3 +12}{{\tilde z}}\right).
 \label{solrs2}
\end{equation}

From a Taylor expansion around $\theta_{1}=0$ we get, $z_{0}=xt^{-\frac{1}{3}}$,
\begin{equation}
{\cal U}^{b}({\tilde z} ,t)={\cal U}^{b}(z_{0},t)+t^{-\frac{1}{3}}\theta_{1}\zeta\dfrac{\partial}{\partial{\tilde z}}{\cal U}^{b}({\tilde z} ,t)\lvert_{{\tilde z}=z_{0}} \label{Taylorexp}
\end{equation}
and since $u^b=u+\theta_1 \xi_1=\partial_x {\cal U}^b$,
we get 
$u(x,t)=t^{-\frac13}  \partial_{z_{0}}{\cal U}^{b}(z_{0},t)$  and $\xi_1=\zeta u_x$, as expected.
The generalization to the infinite set of solutions given in \cite{K,C} is direct and we get
\begin{equation}
{\cal U}^{b}_{n}({\tilde z},t)=i\log\left(t^{\frac{n+1}{3}}\dfrac{Q_{n+1}({\tilde z})}{Q_{n}({\tilde z})}\right),\label{solvb}
\end{equation}
where the functions $Q_{n}({\tilde z})$ are the Yablonskii-Vorob'ev polynomials \cite{C,FOU} define by the recurrence relation
\begin{equation}
 3^{\frac{1}{3}}Q_{n+1}Q_{n-1}={\tilde z}Q_{n}^{2}-12\left(Q_{n}Q_{n;{\tilde z}{\tilde z}}-Q_{n;{\tilde z}}^{2}\right),\label{recYab}
\end{equation}
with $Q_{0}({\tilde z})=3^{-\frac{1}{3}}$ and $Q_{1}({\tilde z})={\tilde z}$. Note that the Yablonskii-Vorob'ev polynomials are used to construct the similarity solutions of the Painlevé II equation \cite{C,FOU}.

 The link with the Hirota formalism is obtained by letting $ \tau_{1}$ and  $\tau_{2}$ be polynomials in the independent variable $\tilde z$ and the time variable $t$. In fact, we take the following series
\begin{equation}
 \tau_{1,n}({\tilde z},t)=t^{\frac{n(n+1)}{6}}Q_{n}({\tilde z}), \ \ \tau_{2,n}({\tilde z},t)=t^{\frac{(n+1)(n+2)}{6}}Q_{n+1}({\tilde z})
\end{equation}
which lead to (\ref{solvb}) and solve the bilinear equations (\ref{H1}) and (\ref{H2}). For example, we see that if $n=1$, we recover the expression (\ref{poly1}).

From the Taylor expansion (\ref{Taylorexp}) and (\ref{solvb}), we get the corresponding $u_{n}(x,t)$ solution of the bosonic superfield $A$, namely
\begin{equation}
 u_{n}(x,t)=it^{-\frac{1}{3}}\dfrac{d}{dz_{0}}\log\left(\dfrac{Q_{n+1}(z_{0})}{Q_{n}(z_{0})}\right), \quad \xi_{1;n}=\zeta f_{1;n}=\zeta\partial_{x}u_{n}.
\end{equation}

Figure 3 shows the behaviour of the imaginary part of $u_{1}$ and $f_{1;1}$ as a function of $x$ and $t$,
\begin{equation}
 u_{1}(x,t)=2i\dfrac{x^{3}-6t}{x(x^{3}+12t)}.
\end{equation}

\begin{figure}[!h]
\begin{center}
 \includegraphics[width=4.5cm]{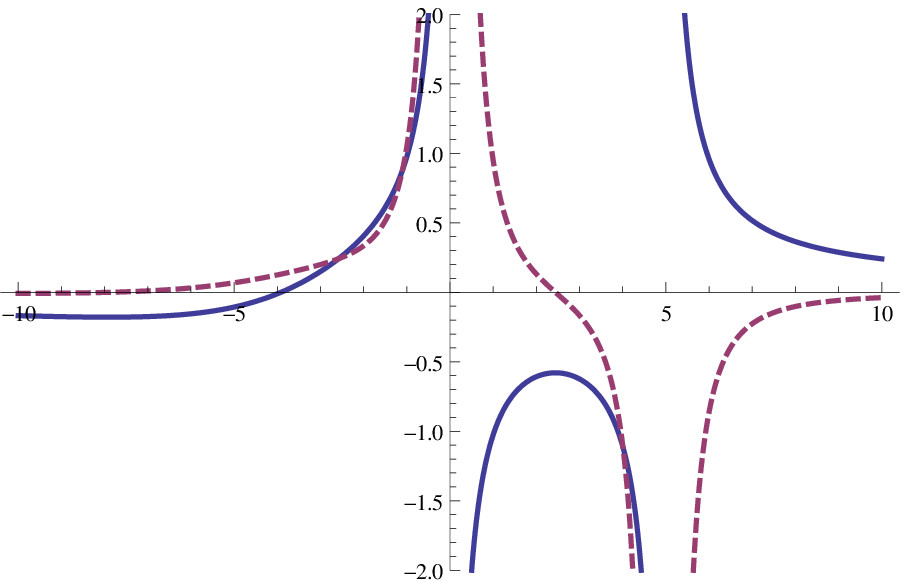}
\includegraphics[width=4.5cm]{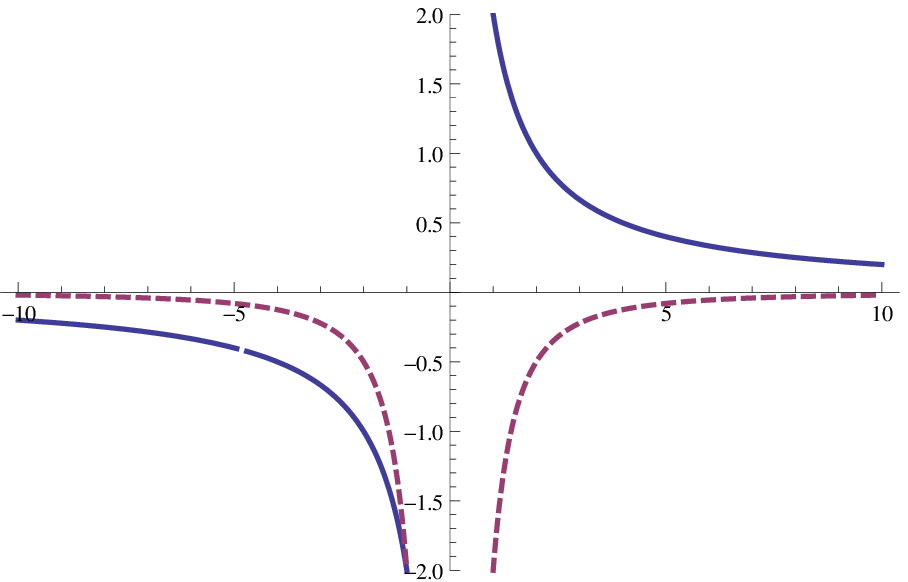}
\includegraphics[width=4.5cm]{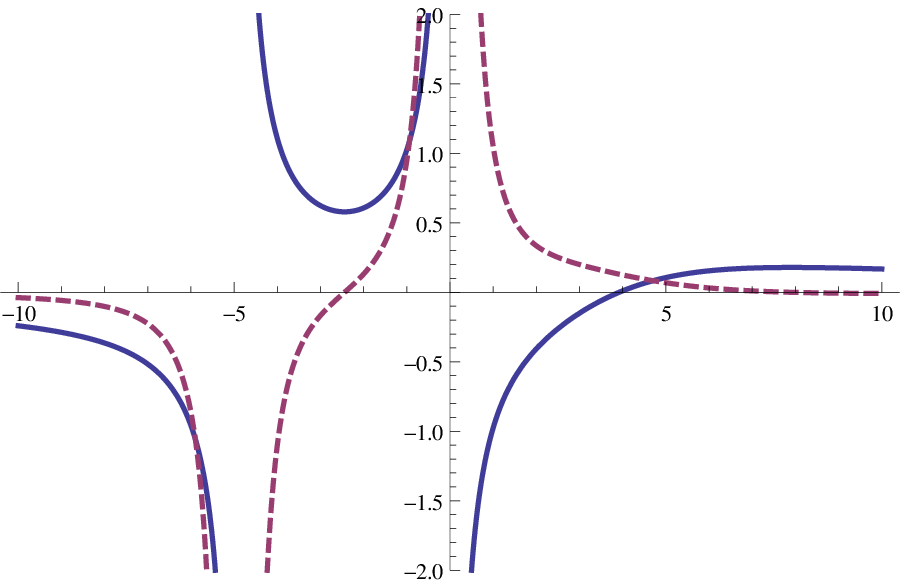}
\caption{The function $Im(u_{1}(x,t))$ (full curves) and $Im(f_{1;1}(x,t))$ (dashed curves) for $t=-10, 0, 10$.}
\end{center}
\end{figure}

\section{Conclusion.}

The generalization of the Hirota formalism to $\mathcal{N}=2$ supersymmetric equations has been used to solve the celebrated  SUSY KdV equation for $a=-2$. We have been able, for the first time, to produce a bilinear form of the $\mathcal{N}=2$ SUSY KdV equation. In fact, the assumption made has produced a reduction of our equation to a $\mathcal{N}=1$ SUSY one for which the bilinearization is known.  From the bilinear equations we have constructed multisoliton solutions and have produced for the first time a SUSY generalization of rational similarity solution. Indeed, we have generated an infinite set of rational solution using a SUSY version of the Yablonskii-Vorob'ev polynomials well known in the construction of similarity solution of the Painlevé II equation.

In future work, we want to use the symmetry reduction method adapted to the SUSY context to create links with the Hirota formalism. This paper suggest such links. Indeed the Yablonskii-Vorob'ev polynomials, which arise from the classical mKdV equation via dilatation invariance, were used in the Hirota formalism to generate an infinite set of solution of the $\mathcal{N}=2$ SUSY KdV equation.

\section*{Acknowledgments.}
L. Delisle acknowledge the support of a FQRNT doctoral research scholarship. V. Hussin acknowledge the support of research grants from NSERC of Canada. 

\section*{References.}

\bibliographystyle{unsrt}

\end{document}